\documentclass[preprint]{aastex}

\slugcomment{Accepted by the ApJ}

\shorttitle{FIRST TASTE OF HOT CHANNEL}

\shortauthors{Song et al.}

\begin{document}
\title{FIRST TASTE OF HOT CHANNEL IN INTERPLANETARY SPACE}
\author{H. Q. SONG\altaffilmark{1}, J. ZHANG\altaffilmark{2}, Y. CHEN\altaffilmark{1}, X. CHENG\altaffilmark{3}, G. LI\altaffilmark{4}, AND Y. M. Wang\altaffilmark{5}}

\affil{1 Shandong Provincial Key Laboratory of Optical Astronomy
and Solar-Terrestrial Environment, and Institute of Space
Sciences, Shandong University, Weihai, Shandong 264209, China}
\email{hqsong@sdu.edu.cn}

\affil{2 School of Physics, Astronomy and Computational Sciences,
George Mason University, Fairfax, VA 22030, USA}

\affil{3 School of Astronomy and Space Science, Nanjing
University, Nanjing, Jiangsu 210093, China}

\affil{4 Department of Space Science and CSPAR, University of
Alabama in Huntsville, Huntsville, AL 35899, USA}

\affil{5 Key Laboratory of Geospace Environment, University of
Science and Technology of China, Chinese Academy of Sciences
(CAS), Hefei, Anhui 230026, China}

\begin{abstract}
Hot channel (HC) is a high temperature ($\sim$10 MK) structure in
the inner corona revealed first by Atmospheric Imaging Assembly
(AIA) on board \textit{Solar Dynamics Observatory}. Eruption of HC
is often associated with flare and coronal mass ejection. Previous
studies suggest that HC is a good proxy of magnetic flux rope
(MFR) in the inner corona, in addition to another well-known MFR
candidate, the prominence-cavity structure that is with a normal
coronal temperature ($\sim$1-2 MK). In this paper, we report a
high temperature structure (HTS, $\sim$1.5 MK) contained in an
interplanetary coronal mass ejection induced by an HC eruption.
According to the observations of bidirectional electrons, high
temperature and density, strong magnetic field, and its
association with the shock, sheath, and plasma pile-up region, we
suggest that the HTS is the interplanetary counterpart of the HC.
\textbf{The scale of the measured HTS is around 14 R$_\odot$}, and
it maintained a much higher temperature than the background solar
wind even at 1 AU. It is significantly different from the typical
magnetic clouds (MCs), which usually have a much lower
temperature. \textbf{Our study suggests that the existence of a
corotating interaction region ahead of the HC formed a magnetic
container to inhibit the HC expansion and cooling down to a low
temperature.}
\end{abstract}


\section{INTRODUCTION}
Hot channel (HC) refers to the high temperature structure that is
revealed first by coronal images of AIA (Atmospheric Imaging
Assembly) 131 \AA \ passband (sensitive to temperature of $\sim$
10 MK), while the structure is invisible from cooler temperature
images, e.g., images of the AIA 171 \AA \ passband (sensitive to
temperature of $\sim$ 0.6 MK) (Zhang et al. 2012; Cheng et al.
2013a, 2013b, 2014a, 2014b, 2014c; Li \& Zhang 2013). HC appears
as a hot blob structure if observed along the channel axis (Cheng
et al. 2011; Patsourakos et al. 2013; Song et al. 2014a, 2014b)
due to the projection effect. Hereafter, we will use HC to refer
to both hot channel and hot blob structures.

HC has been generally regarded as a proxy of magnetic flux rope
(MFR, a volumetric plasma structure with the magnetic field lines
wrapping around a central axis) since its discovery with AIA on
board \textit{Solar Dynamics Observatory (SDO)}. This is supported
by the following observational studies: (1) Cheng et al. (2014a)
observed an HC that showed helical threads winding around an axis.
In the meantime, cool filamentary materials descended spirally
down to the chromosphere, providing direct observational evidence
of intrinsical helical structure of HC; (2) Cheng et al. (2011)
reported that HC can grow during the eruption, similar to the MFR
growth process according to the classical magnetic reconnection
scenario in eruptive flares; Song et al. (2014a) presented the
formation process of an HC during a CME and found that the HC was
formed from coronal arcades through magnetic reconnection. Their
works further support that the HC is an MFR structure based on the
relation between HC and magnetic reconnection; (3) Cheng et al.
(2014b) found an HC was initially cospatial with a prominence,
then a separation of the HC top from that of the prominence was
observed during the eruption initiated by the ideal kink
instability (T\" or\" ok et al. 2004). It is widely accepted that
prominence/filament can exist at the dip of a flux rope (Rust \&
Kumar 1994). Therefore, this observation offered another important
support that HC is an MFR;

Except HC, several lines of observations in the lower corona have
also been proposed as MFRs, including sigmoid structure in active
region (Titov \& D\'{e}moulin 1999; Mckenzie \& Canfield 2008) and
coronal cavity in quiescent region (Wang \& Stenborg 2010). A
sigmoid has either a forward or reverse S-shape with enhanced
X-ray emissions (implying an entity of high temperature) with its
center straddling along the polarity inversion line of the hosting
active region. Zhang et al. (2012) showed that the HC initially
appeared like a sigmoidal structure and then changed to a
semi-circular shape. Therefore, sigmoid and HC might represent the
same structure, their different shapes are likely from different
perspectives and evolution phase. Both structures are featured by
high temperature, a possible result of flare magnetic reconnection
(e.g., Song et al. 2014a, 2014b). Coronal cavity, on the other
hand, observed as dark circular or oval structure above solar limb
in coronal images with temperatures close to the background corona
(Fuller et al. 2008; Gibson et al. 2010; Kucera et al. 2012), is
also interpreted as MFR. As mentioned, the long-studied feature of
solar filament/prominence shown best in H$\alpha$ images has been
interpreted as situated along the dip in MFR. Therefore,
prominence lying in the dip of coronal cavity is not rare. The
eruption of coronal cavity (or filament) from quiescent region
doesn't show high-temperature signature like HC, which might be
attributed to lack of obvious heating acquired from the weak
magnetic reconnection (e.g., Song et al. 2013).

According to the descriptions above, at least two different types
of MFRs can be identified in the inner corona depending on their
temperatures, i.e., high-temperature MFR like HC and
low-temperature MFR like coronal cavity. Note that it is possible
that the HC has a low initial temperature but heated later by
flare magnetic reconnection during the eruption (e.g., Song et al.
2014a, 2014b). One obvious question arises as what the difference
is between these two MFR structures when they are detected in situ
near 1 AU. Magnetic cloud (MC), with lower temperature than the
background solar wind, is a well known interplanetary structure
(Burlaga et al. 1981; Lepping et al. 1990). Can the HC maintain
its higher temperature than the background at 1 AU, or will it
evolve into a cool MC? In this paper, we will try to address this
question with instruments on board \textit{Solar TErrestrial
RElations Observatory (STEREO)} through tracing an HC eruption
from the Sun to $\sim$ 1 AU. In section 2, we introduce the
instruments. The observations and discussion are presented in
Section 3, which are followed by a summary in our last Section.

\section{INSTRUMENTS}
Our event was observed by three spacecraft including \textit{SDO},
\textit{SOHO (Solar and Heliospheric Observatory)}, and
\textit{STEREO}. The AIA on board \textit{SDO} provides the solar
atmosphere images in 10 narrow UV and EUV passbands with a high
cadence (12 seconds), high spatial resolution (1.2 arcseconds) and
large FOV (1.3 R$_\odot$). The AIA passbands cover a large
temperature range from 0.6 to 20 MK (O'Dwyer et al. 2010; Del
Zanna et al. 2011; Lemen et al. 2012). During an eruption, the
131~\AA\ passband is sensitive to the hot plasma from flare
regions and erupting HC (e.g., Zhang et al. 2012; Cheng et al.
2011; Song et al. 2014a, 2014b). AIA's high cadence and broad
temperature coverage make it possible for constructing
differential emission measure (DEM) models of corona plasma (Cheng
et al. 2012 and references therein). In addition, the COR
coronagraph instrument (Howard et al. 2008) on board
\textit{STEREO} (Kaiser et al. 2008) and LASCO on board
\textit{SOHO} (Domingo et al. 1995) provide CME images in the
outer corona from different perspectives. Heliospheric Imager (HI,
Howard et al. 2008) on board \textit{STEREO} images the whole
propagation process of the associated ICME from near the Sun to
$\sim$ 1 AU. PLASTIC and IMPACT on board \textit{STEREO} measure
the solar wind properties and interplanetary magnetic field. Data
from the above instruments are analyzed in the following section.

\section{OBSERVATIONS AND DISCUSSION}
On 2012 January 27, an X1.7 class soft X-ray (SXR) flare was
recorded by the \textit{Geostationary Operational Environmental
Satellite (GOES)}, which started at 17:37 UT and peaked at 18:37
UT. The flare location was at $\sim$N33W85 (NOAA 11402) from the
perspective of the Earth. Figure 1 shows the positions of
different spacecraft in the ecliptic plane, including
\textit{SDO/SOHO}, \textit{STEREO} A and B. During this flare,
\textit{STEREO} A and B were 107.8$^{\circ}$ west and
114.5$^{\circ}$ east of the Earth with a distance of 0.96 AU and
1.06 AU, respectively. Therefore, the source location on the Sun
was $\sim$23$^{\circ}$ east of the central meridian as viewed from
\textit{STEREO} A, whereas $\sim$70$^{\circ}$ behind the west limb
for \textit{STEREO} B. Obviously, \textit{STEREO} A provides the
best disk observation of the active region, while \textit{SDO} and
\textit{SOHO} give the limb views of the eruption.

\subsection{HC Eruption in the Inner Corona}
For this event, a very clear HC can be observed during the
eruption, rising from 17:37 UT onward and arriving at the rim of
AIA FOV at 18:15 UT. The HC showed an interesting morphological
evolution from a channel with twisted or writhed axis (Figure
2(a)) to a channel with loop-like axis (Figure 2(c)), as indicated
by the dotted lines. This morphological evolution is very similar
to the event reported by Zhang et al. (2012). During the
evolution, the two footpoints of the evolving HC remained fixed on
the Sun (see the first animation accompanying Figure 2 for the
whole process). To describe the overall thermal properties of the
HC, DEM-weighted temperature maps (see Cheng et al. 2012 and Song
et al. 2014b for the validation and other details) are
reconstructed and presented in Figures 2(b) and (d), which show
the HC temperature is around 10 MK at the times of Figures 2(a)
and (c), respectively. \textbf{Here we also acquire the HC density
through DEM analysis (see Cheng et al. 2012 for the method), which
is around 10$^{9}$ cm$^{-3}$ and much higher than the density of
its surrounding corona at the same altitude.} By carefully
inspecting the AIA and LASCO animations, one can deduce that the
HC eruption induced a CME (see the second animation accompanying
Figure 2), which was recorded by LASCO and COR from three distinct
perspectives as described in the following subsection. With
combined observations of \textit{SDO}, \textit{STEREO} A and B, we
conclude that no other CMEs or large blowout jets took place
during the time of interest (see the third animation accompanying
Figure 2), which concludes that the CME was caused by the HC
eruption.

\subsection{CME Observations in the Outer Corona}
In the outer corona, the CME was well observed by the LASCO, COR-A
and COR-B instruments as shown in Figures 3(a)-(c) (also see the
accompanying animation). The CME appeared in LASCO C2 FOV first at
18:27 UT, and its linear speed was 2508 km s$^{-1}$ in the LASCO
C2/C3 FOV. The three viewpoints provide three distinct projections
of the CME. We can distinguish a coherent bright structure and a
preceding CME front region in all three perspectives. The CME
front region ahead of the MFR likely consists of three components:
plasma pile-up of the MFR, an outer diffuse shock front and the
sheath region between them (Vourlidas et al. 2013; Cheng et al.
2014a). Through inspecting the HC eruption and CME propagation in
LASCO FOV carefully, we believe that the coherent bright structure
and preceding front region are the HC and pile-up plasma,
respectively, which is consistent with the conclusions of Cheng et
al. (2014a). This is further supported by the graduated
cylindrical shell (GCS) model (Thernisien et al. 2006)

Using the GCS model of Thernisien et al. (2006), we can
reconstruct the three-dimensional (3D) morphology of the HC. The
model depends on six parameters: the source Carrington longitude
($\phi$) and latitude ($\theta$), the MFR tilt angle ($\gamma$),
height ($r$) and aspect ratio ($\kappa$), as well as the
half-angle ($\alpha$) between the two legs of MFR. We first
estimate $\phi$ (186$^{\circ}$), $\theta$ (37$^{\circ}$), and
$\gamma$ (79$^{\circ}$) using the location and neutral line of the
active region through the Extreme Ultraviolet Imager (EUVI) 195
\AA \ images, then vary $\alpha$ (57$^{\circ}$), $\kappa$ (0.17),
and $r$ (5.6 R$_\odot$) till we achieve the best visual fit in the
three coronagraph images simultaneously. The numbers in the
brackets are the final positioning and model parameters of the HC
for the time shown in Figure 3. The results are displayed in
Figures 3(d)-(f). It's clear that LASCO and COR-A were observing
the HC face on, and COR-B edge on. Therefore, the HC appeared as a
bright channel in LASCO and COR-A FOV and a bright blob in COR-B
FOV. It's clear that our CME is a limb event from the Earth
perspective, and the HC is almost along the west solar limb. With
the fitting results of GCS model and assuming that the HC
experienced a self-similar expansion (M\"ostl et al. 2014), we got
the longitude range of the HC is not over 40 degrees, which was
shown with red dash lines in Figure 1, if assuming the CME
propagated outward radially in the ecliptic plane along the red
solid line in Figure 1. However, we note that the CME might
deflect in the corona and interplanetary space (Wang et al. 2004,
2013; Gopalswamy et al. 2009; Shen et al. 2011; Gui et al. 2011).
Figure 1 shows that the MFR will be likely detected by
\textit{STEREO A}, with the spacecraft trajectory far away from
its center, which might influence the in-situ detection of the MFR
(D\'{e}moulin et al. 2013; Riley \& Richardson 2013). The in-situ
observations will be discussed in Section 3.4.

It's well accepted that the typical morphology of a normal CME
contains the so-called three-part structure: a bright front loop,
a dark cavity and an embedded bright core (Illing \& Hundhausen,
1985), corresponding to the pile-up plasma, MFR, and the erupting
filament (House et al. 1981), respectively. However, for CME
induced by an HC eruption without a filament, the embedded bright
part corresponds to the HC, instead of the filament. In this case,
the CME will show a bright front loop and a coherent bright
structure, corresponding to the pile-up plasma and HC (or MFR),
respectively. It's reasonable because the HC is not only hotter,
but also denser than the background plasma (Cheng et al. 2012).
The shock can be generated if CMEs move fast enough. In our event,
the shock, pile-up plasma, and HC (MFR) can be observed directly
in the coronagraphic FOV as depicted with arrows in Figure 3(c).
Usually, the diffuse front ahead of the pile-up region is
interpreted as a shock structure (e.g., Vourlidas et al. 2003,
2013; Feng et al. 2012, 2013), and the diffusive layer corresponds
to the sheath region. A type II solar radio burst associated with
this event was detected (not shown here), which further confirmed
the existence of a shock. Therefore, in this event we expect that
the shock, sheath, pile-up plasma (front region), HC (MFR), and
remainder of the ICME (rear region) are all observed by the
coronagraphs, and may have their corresponding in-situ
counterparts (e.g., Kilpua et al. 2013), as will be presented
later.

\subsection{ICME Propagation in Interplanetary Space}
The CME propagation in interplanetary space was well observed by
HI-1 and HI-2, as presented in Figures 4(a) and (b). The ICME
first appeared in the HI-1A FOV at 19:29 UT on January 27, and in
the HI-2A FOV at 02:09 UT on January 28. We produce a
time-elongation map by stacking the running difference images
within a slit along the ecliptic plane as shown in Figures 4(a)
and (b) with the red rectangle, and present it in Figure 4(c).
Here to trace the propagation of ICME in interplanetary space, we
just use HI-1 and HI-2 images. Note that the elongation angles are
plotted in a logarithmic scale to expand HI-1 data, so tracks are
not J-like as in traditional linear-linear plots (Liu et al.
2010). The time-elongation map shows one obvious and continuous
track as indicated with the red dotted line. The vertical red line
in Figure 4(c) depicts the arrival time of the ICME shock to
\textit{STEREO} A, which is 13:04 UT on January 29. And no other
ICME propagation was observed by HI from near the Sun to $\sim$1
AU during these days.(see the animation accompanying Figure 4 for
the whole propagation process). These observations show that the
ICME detected by \textit{STEREO} A is the one we are tracing.

\subsection{ICME (HC) Detection near 1 AU}
Figure 5 shows the in situ measurements from the IMPACT and
PLASTIC instruments on board \textit{STEREO} A at 0.96 AU. From
top to bottom, the panels show the normalized pitch angle (PA)
distribution of 93.47 eV electrons (with electron flux values
descending from red to black), the proton bulk speed (black line)
and ratios of three components to the total speed, magnetic field
strength (black line) and its three components, proton density and
temperature, plasma $\beta$ and total pressure, and entropy. Note
the velocity (panel b) and magnetic field (panel c) components are
plotted in RTN coordinates, where R (red line) points from the Sun
center to the spacecraft, T (green line) is parallel to the solar
equatorial plane and along the direction of planet motion and N
(blue line) completes the right-handed system.

As mentioned in Section 3.2, we expect that the shock, sheath,
pile-up plasma, HC (MFR), and remainder of ICME can be detected
one by one with in situ measurements. An obvious forward shock
(depicted with 1 in panel b) passed \textit{STEREO} A at 13:04 UT
on January 29. The transit time is 43.5 h taking the flare start
time (17:37 UT on January 27) to be the CME launch time. One ICME
can be identified from the magnetic field data behind the shock.
The PA distributions in panel a distinguish the different parts of
ICME. The sheath region is very turbulent (e.g., Burlaga et al.
1981), so electrons presented PA between 0 $\sim$ 180$^{\circ}$ in
this region (depicted with 2 in panel b, the left shaded region),
while for the pile-up region, the anti-parallel electron flow
dominated (depicted with 3 in panel b, between the two shaded
regions), similar to the background solar wind, supporting that it
is the pile-up materials of background plasma. Bidirectional
electrons (BDEs) appeared within a high-temperature structure
(HTS, $\sim$1.5 MK, as depicted with 4 in panel b in the right
shaded region), indicating that it corresponds to a magnetic
structure with both footpoints anchored on the Sun. The remainder
of ICME is depicted with 5 in panel b. The final part likely ends
around 18:00 UT on January 30 as indicated with the vertical blue
dot dash line, when the magnetic filed, temperature, and total
pressure approach to the background values.

\subsection{Discussion}

The total magnetic field strengths in the shock sheath and HTS
keep around $\sim$45 nT and $\sim$20 nT, respectively, and vary
between 30 and 50 nT in plasma pile-up region. The R and T
components of HTS keep almost constant while the N component
direction shows irregular rotation, which will be explained later.
The density of HTS is $\sim$15 cm$^{-3}$ and higher than the
background solar wind, while it's lower than that of the sheath
and plasma pile-up region (panel e) due to its expansion during
propagation from near the Sun to $\sim$1 AU. Based on its BDEs,
high temperature, strong magnetic field strength, high density,
and its association with the shock, sheath, and plasma pile-up
region, we suggest that the HTS is the interplanetary counterpart
of the HC observed in lower corona as shown in Figure 2. The
presence of the embedding high Fe charge state further supports
this conclusion, which will be discussed later. The HC started at
19:00 UT and ended at 23:50 UT, the average bulk velocity is 570
km s$^{-1}$ during this period (panel b), so the scale of the
measured HC is around 14 R$_\odot$. The plasma $\beta$ in the HC
is around 1 (panel e), which means the thermal pressure is nearly
equal to the magnetic pressure. The high thermal pressure is
attributed to the high temperature. The entropy in the HC region
is considerably higher than its surroundings (panel f). From above
descriptions, we find the temperature and density of HC decreased
from $\sim$10 MK and $\sim$10$^{9}$ cm$^{-3}$ to $\sim$1.5 MK and
$\sim$15 cm$^{-3}$ from near the Sun to $\sim$1 AU, respectively.

According to the ICME list provided on the \textit{STEREO}
website\footnote{$http://stereo-ssc.nascom.nasa.gov/data/ins_data/impact/level3/$},
this ICME is sorted into Group 3, which means the spacecraft
passed far away from the ICME center, displaying a rapid rise and
then gradual decay in total pressure (Jian et al. 2006). It is
consistent with our CME propagation analysis in Figure 1. This may
lead to two consequences as mentioned above: First, the scale of
the measured HC is small compared to the typical MC structure near
1 AU, which is around 0.25 AU (over 50 R$_\odot$) (see, e.g.,
Lepping et al. 2006); Second, it is not easy to observe a regular
rotation of magnetic field. Therefore, we do not acquire a nice
MFR structure with the Grad-Shafranov (GS) reconstruction method
(Hu \& Sonnerup 2002), which works best for spacecraft passing
near the ICME center. The weakening of the MFR signature with
increasing distance of the spacecraft from the ICME center has
been demonstrated by multi-spacecraft observations (Cane et al.,
1997; Kilpua et al. 2011), consistent with our observations.

As mentioned above, an MC (Burlaga et al. 1981) can be frequently
identified in ICME structures, usually behind the shock, sheath,
and plasma pile-up region. The magnetic field vectors in a typical
MC are observed to have a large rotation, consistent with the
passage of an MFR. The field strength is high, and the density and
temperature are relatively low with a low plasma $\beta$ (less
than 0.1, see Lepping et al. 1997). The total pressure inside the
cloud is higher than outside, causing the cloud to expand with its
propagation, even to a distance beyond 1 AU (Burlaga et al. 1981).
However, in our case, an ICME structure with a much higher
temperature ($\sim$1.5 MK) and irregular rotation of Bn was
detected, and the associated plasma $\beta$ was around 1, which
obviously is not the traditional MC. \textbf{According to a very
recent statistical study based on 325 ICMEs from 1996 to 2008
(Mitsakou \& Moussas 2014), the temperatures of ICMEs at 1 AU are
usually lower than 0.25 MK, and their averaged value is only 0.076
MK.} We conjecture that there exist two types of interplanetary
MFR (IMFR) structures mainly according to their temperatures,
i.e., the low-temperature IMFR (or MC) corresponding to MFR (e.g.,
coronal cavity) without obvious heating during its eruption (e.g.,
Song et al. 2013), and the high-temperature IMFR corresponding to
MFR (e.g., HC) with significant heating during or before its
eruption (e.g., Song et al. 2014a, 2014b). In our event, the later
can keep its temperature higher than background even to 1 AU. It
might be confusing why the temperature of HC didn't decrease to a
level lower than the background wind through its faster expansion
in the interplanetary space. To address this, we note that the
total pressure ahead of the HC is much higher (see Figure 5(e))
than the usual solar wind, which might prevent the HC from a free
expansion.

According to the statistical study (Richardson \& Cane 2010; Wu \&
Lepping 2011), MCs are detected in only about 30\% of ICMEs. Riley
and Richardson (2013) listed several explanations for why some
ICMEs are observed to be MCs and others are not, e.g., the
observational selection effect of ICMEs, the interactions of an
MFR with itself or between neighboring MFRs, the effect of
evolutionary process of MFRs, and the different initiation
mechanisms of CMEs. As mentioned above, there are different
observational lines raised as proxies of MFRs in the lower corona,
e.g., filaments/prominences, coronal cavities, sigmoid structures,
and hot channels. Therefore, it's natural to argue that ICMEs with
or without MCs might correspond to different coronal structure
eruptions. Our results indicate that the HC eruption might not
evolve into a typical MC under some special conditions. More
events are necessary to conclude this point.

If the HTS really corresponds to HC in the lower corona, then we
should be able to detect high charge state of Fe element with in
situ measurements, because the charge state distribution is fully
established within a few solar radii from the Sun, and remains
frozen in after that (e.g., Esser \& Edgar, 2001; Chen et al.
2004). Unfortunately, high temporal resolution Fe charge state
data is not available for this event. The ICME list provided on
\textit{STEREO} website (the same address with above) indicated
that there was a significant increase of Fe charge state during
our event, which hints the coronal origin of the HTS and supports
our conclusion.

It should be mentioned that a weak shock was observed at 2:13 UT
on January 29 before the ICME shock (See the red arrow in Figure
5(b)). It seems to be a forward shock generated by a corotating
interaction region (CIR, see e.g., Wu et al. 2014), \textbf{whose
presence is supported by the appearance of a low latitude coronal
hole ahead of NOAA active region 11402 according to the
observations of the X-ray telescope on board \textit{HINODE}}. As
mentioned, this CIR structure is the reason for the presence of
the high-pressure region ahead of the HC, \textbf{which acts as an
obstacle and inhibits the HC expansion. We suggest that a
preceding CIR (or ICME, e.g., Liu et al. 2014) shall be a
necessary condition for the presence of a HC at 1 AU.} It is
likely that the CME-driven shock ran into the CIR, which makes the
interplanetary transient looks complex as presented in Figure 5.
Regions 2 and 3 in Figure 5 might include the compressed CIR
plasma. Nevertheless, we believe that the ICME-CIR interaction
will not change our interpretation of the detected HTS based on
the descriptions and discussion of BDEs, magnetic field,
temperature, and total pressure. As mentioned, the different
trajectories of spacecraft through ICME make the observational
characteristics of ICME difference. For this event, it also seems
that the regions 2, 3, and 4 are all belong to the sheath, and
just region 5 corresponds to the ejecta according to Figure 5(b).
However, we think this possibility is not high because the total
magnetic field in region 5 is at the background level, and the
BDEs analysis in Figure 5(a) doesn't support this point, either.

\section{SUMMARY}
In this paper, an HC eruption associated with an X1.7 class SXR
flare was recorded by \textit{SDO} and \textit{GOES}. The
corresponding fast CME can be well observed from three distinct
viewpoints by coronagraphs on board \textit{SOHO}, \textit{STEREO}
A and B. The shock, pile-up region and HC can be well observed in
coronagraphic FOVs. And the HC (coherent bright structure) in
coronagraph images can be well fitted with the GCS model. The CME
propagation into the interplanetary space can be traced with the
HI-1/2 instruments, and detected in-situ by instruments on board
\textit{STEREO} A. Further, no other ICME propagation in HI FOV
during these days. This concludes that the HI ICME is the HC
eruption we are tracing. For the first time, we might taste the HC
in interplanetary space, which is mainly identified by its high
temperature, appearance behind shock, sheath and pile-up region,
and the BDEs. The preliminary Fe charge-state report from the
\textit{STEREO} team further supports that the high temperature
property observed near 1 AU has its origin in the inner corona.
Compared with the background solar wind, the interplanetary HC has
a strong magnetic field, and shows obvious BDE flow, indicating
its two footpoints still connecting to the Sun. This supports that
the interplanetary HC belongs to an MFR structure. Nevertheless,
it's likely that the spacecraft passed far away from the ICME
center, so the rotation of magnetic field components was not
obvious and it's difficult to obtain a nice flux rope structure
with the GS reconstruction method. In future studies, we expect
that a suitable event will enable us to observe the known MFR
signatures in the aftermath of a HC eruption.

\acknowledgments We thank the referee for constructive comments
that have greatly improved this manuscript. We are grateful to L.
Jian, B. Li, Q. Hu, Q. M. Lu, C. L. Shen and C. L. Tang for their
valuable discussions. \textit{SDO} is a mission of NASA's Living
With a Star Program, \textit{SOHO} is a mission of international
cooperation between ESA and NASA, and \textit{STEREO} is the third
mission in NASA's Solar Terrestrial Probes program. This research
is supported by the 973 program 2012CB825601, NNSFC grants
41274177, 41274175, and 41331068. J. Zhang is supported by NSF
grant ATM-0748003, AGS-1156120 and AGS-1249270. G. Li is supported
by ATM-0847719 and AGS-1135432.

\clearpage

\begin{figure}
\epsscale{0.8} \plotone{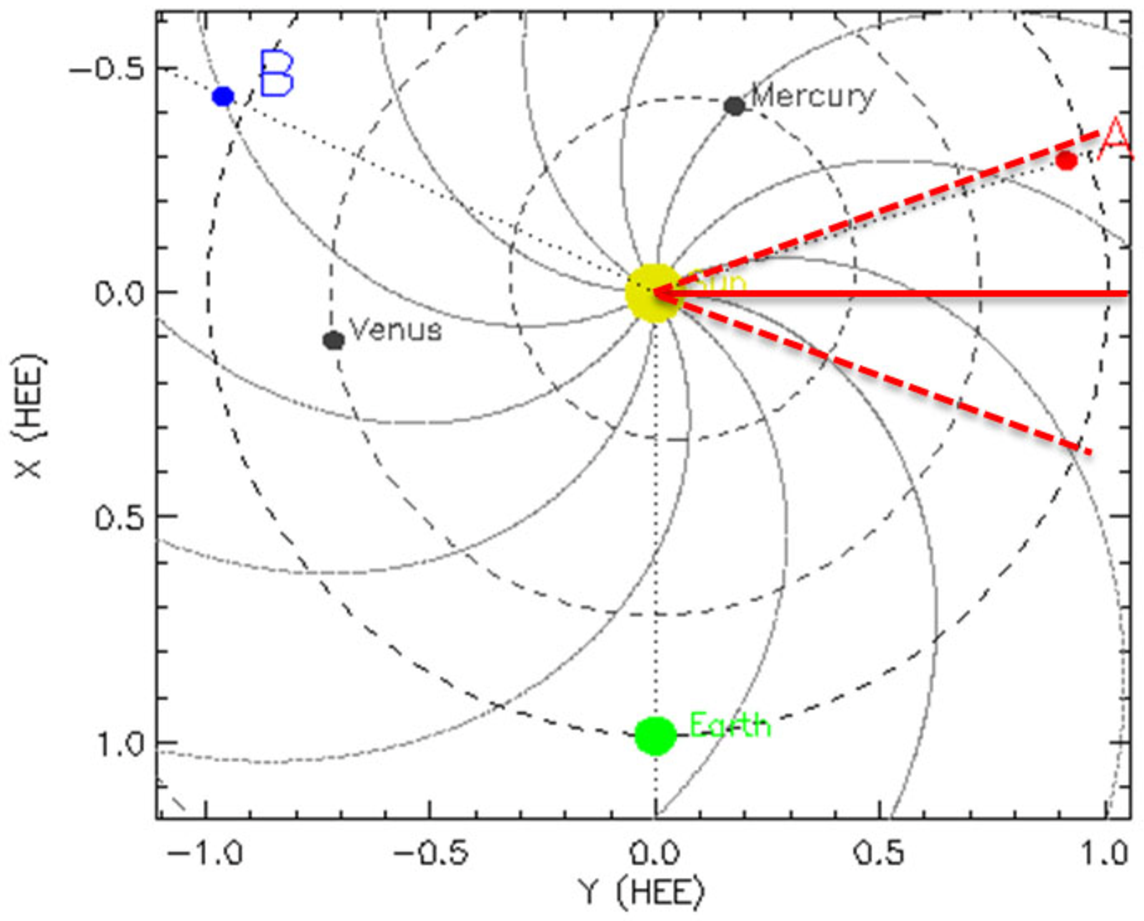} \caption{Positions of the
spacecraft and planets, including the Parker spiral magnetic field
lines, in the ecliptic plane on 2012 January 27. The dashed
circles indicate the orbits of the Mercury, Venus, and Earth. The
dotted lines show the spiral interplanetary magnetic fields. The
radial trajectory of the CME in the ecliptic plane is depicted by
red solid line, and the red dash lines indicate the longitude
range of MFR propagation outward. (A color version of this figure
is available in the online journal.) \label{fig1}}
\end{figure}

\begin{figure}
\epsscale{0.8} \plotone{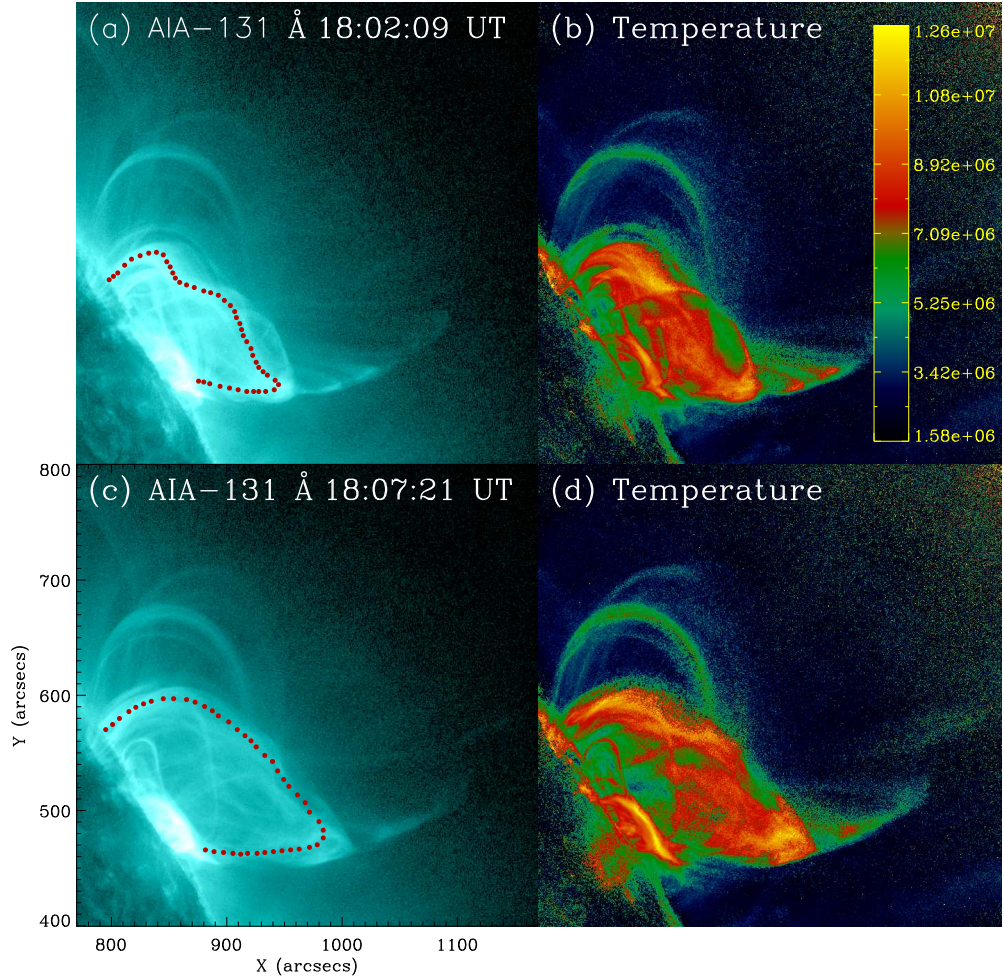} \caption{The HC eruption process
on 2012 January 27. (a), (c) AIA 131 \AA \ image. (b), (d)
Temperature images deduced with the DEM method. (Animations and a
color version of this figure are available in the online
journal.)\label{fig2}}
\end{figure}

\begin{figure}
 \epsscale{0.32} \plotone{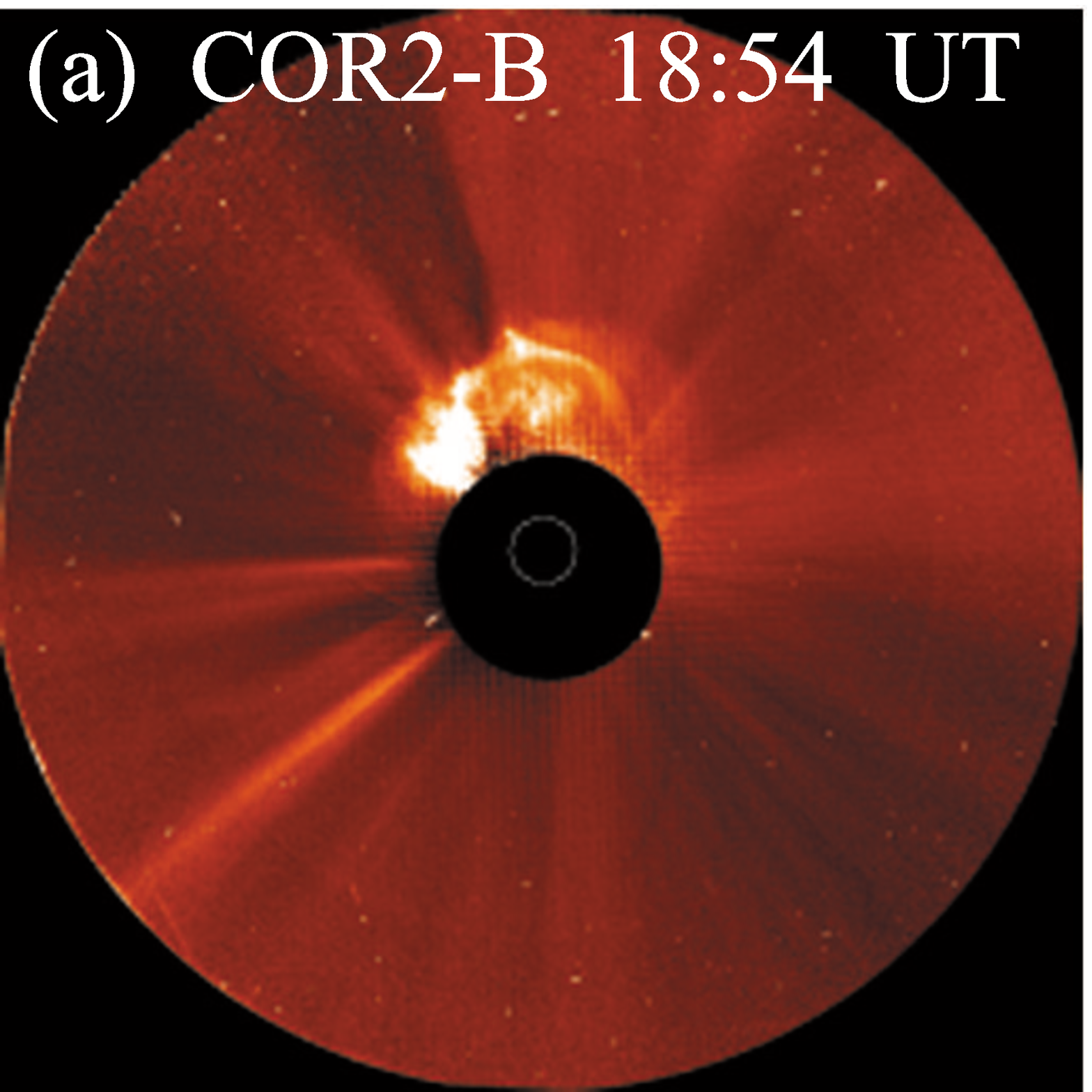}
 \epsscale{0.32} \plotone{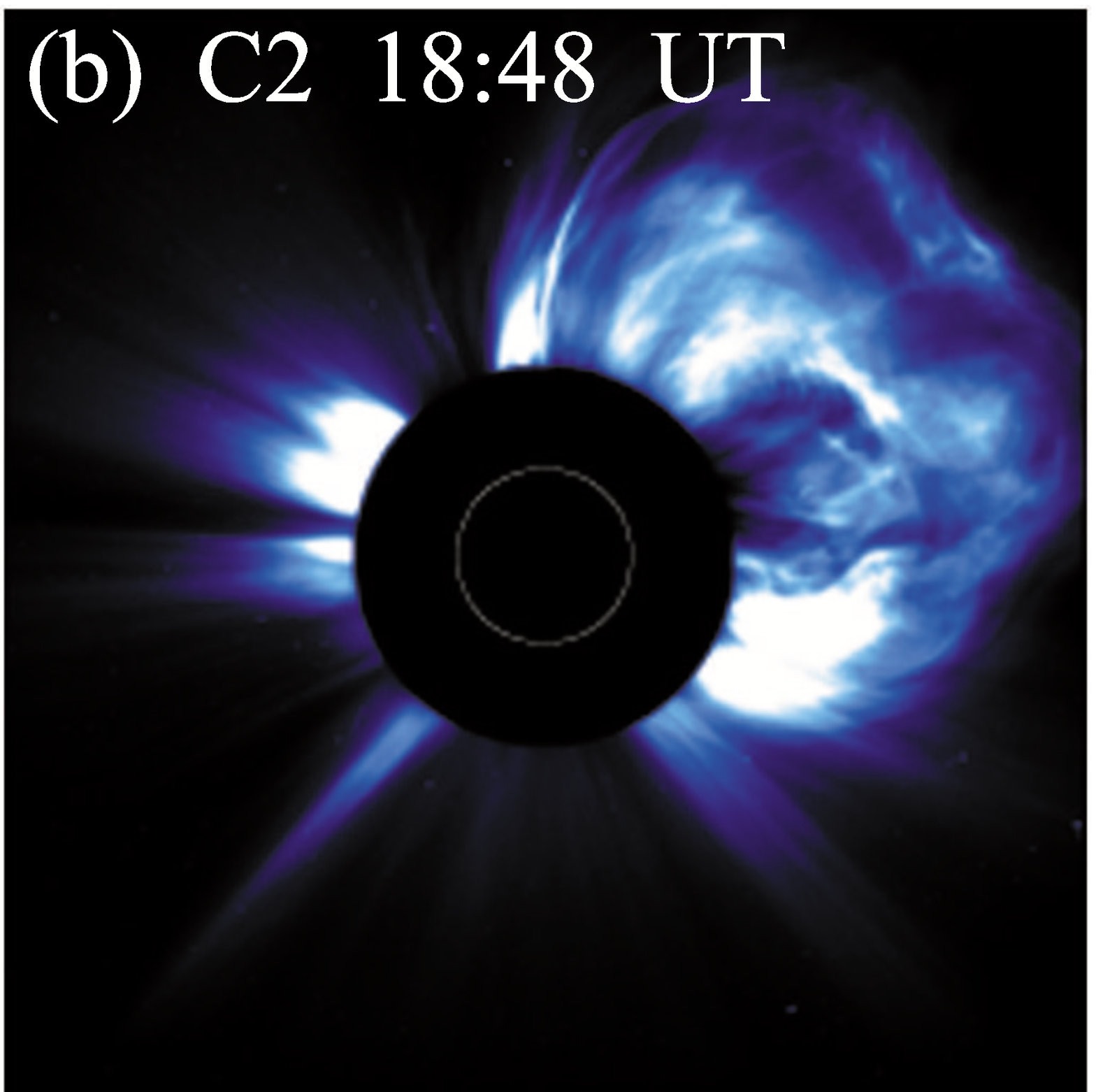}
 \epsscale{0.32} \plotone{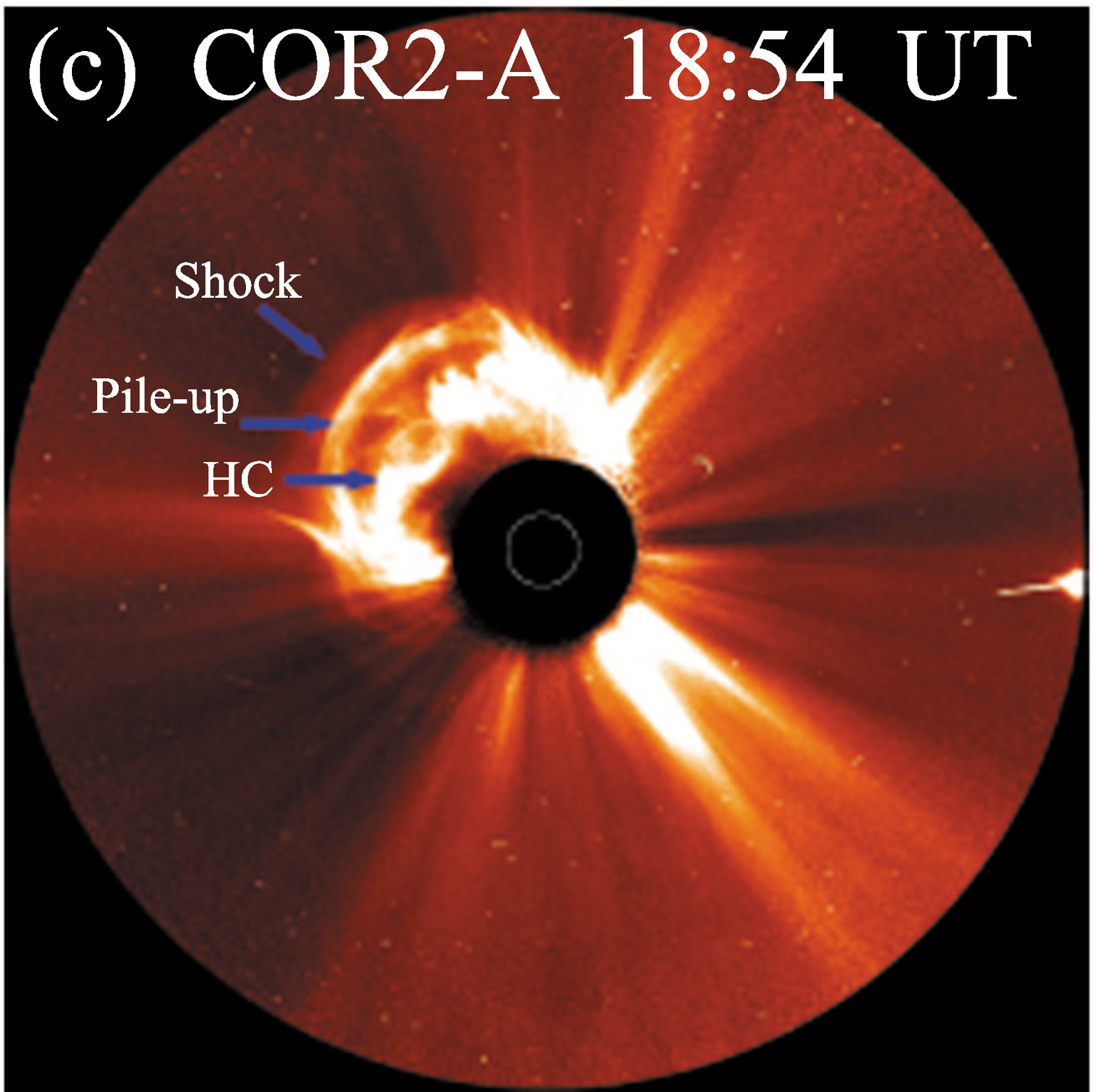}
 \epsscale{0.32} \plotone{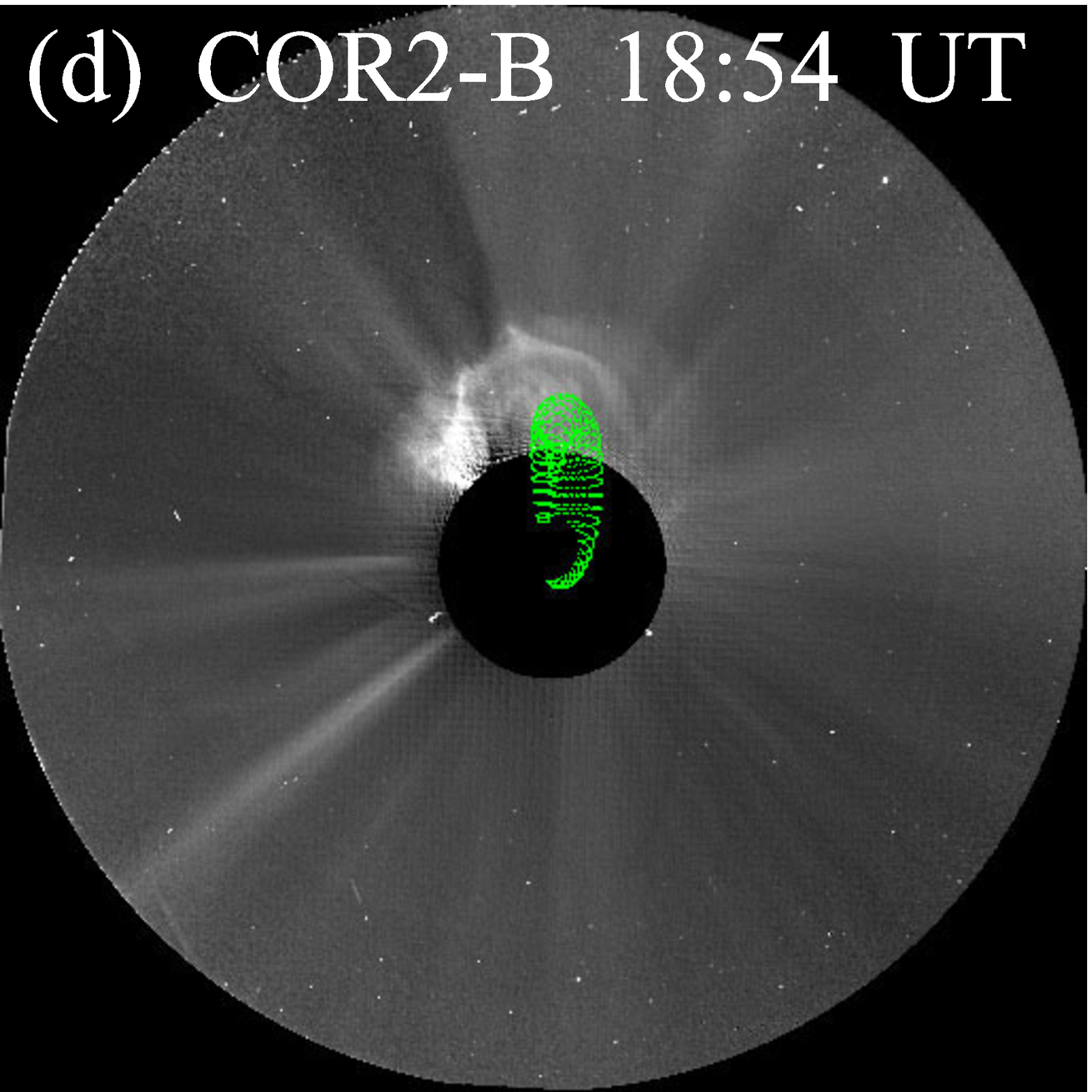}
 \epsscale{0.32} \plotone{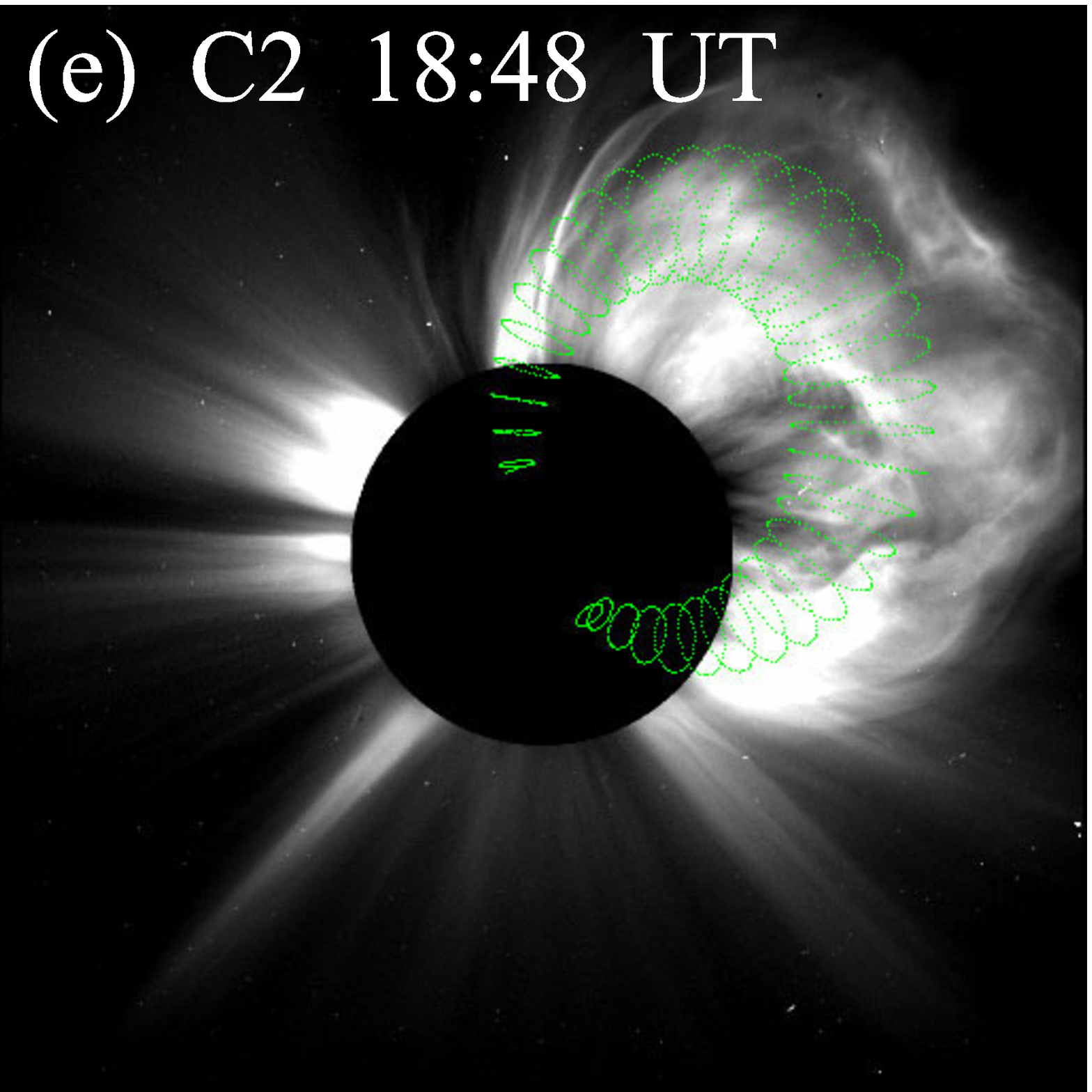}
 \epsscale{0.32} \plotone{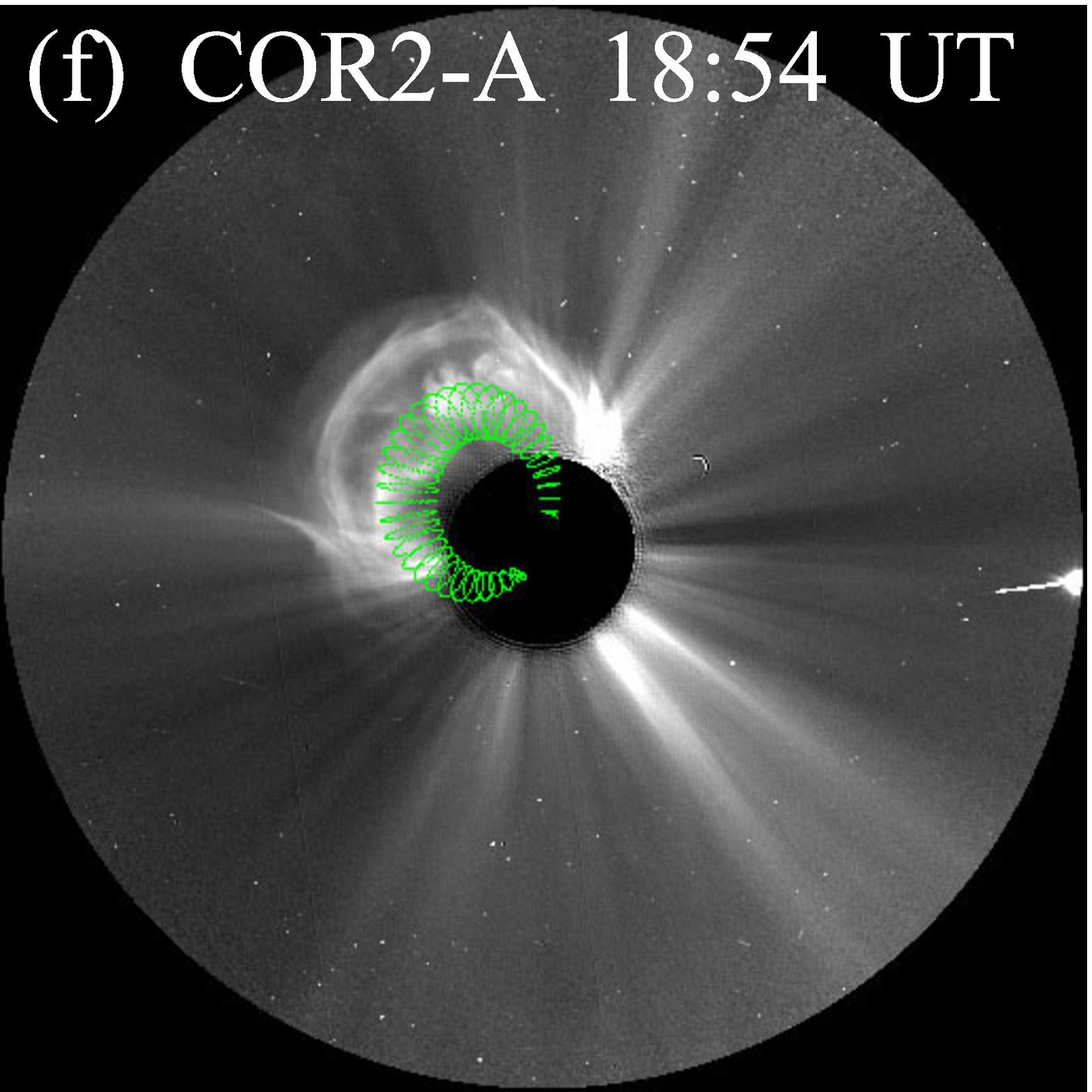}
 \caption{(a)-(c) COR2 and LASCO/C2 white-light coronagraph
images of the eruption. The white circles denote the location of
the solar limb, and the black disks are the coronagraph blocking
plates. (d)-(f) White-light coronagraph images with GCS
reconstruction results of the HC/MFR (green lines) superposed. (An
animation and a color version of this figure are available in the
online journal.)\label{fig3}}
\end{figure}

\begin{figure}
\epsscale{0.8} \plotone{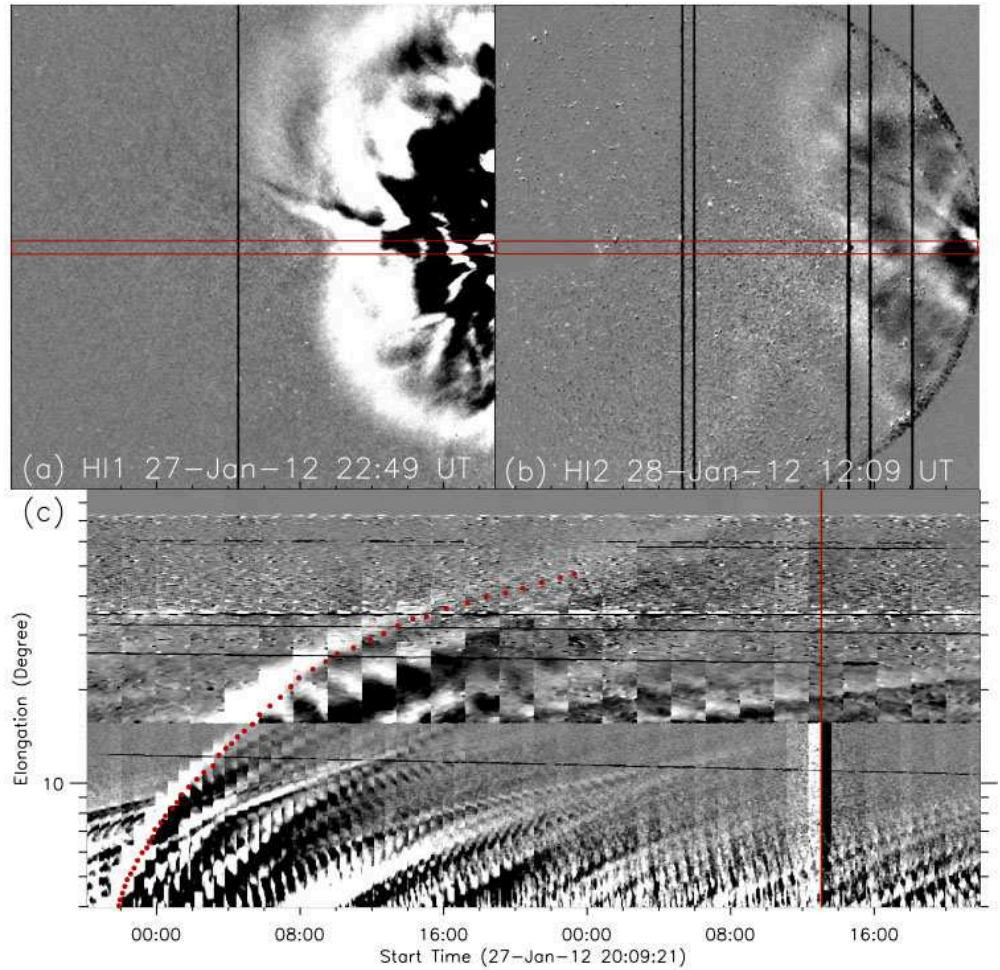} \caption{ICME propagation in
interplanetary space. (a), (b) HI-1 and HI-2 observations of the
ICME, respectively. (c) Time-elongation maps constructed from
running difference images along the ecliptic, as indicated with
the red rectangles in (a) and (b). The vertical red line indicates
the arrival time of ICME shock to \textit{STEREO} A. (An animation
and a color version of this figure are available in the online
journal.)\label{fig4}}
\end{figure}

\begin{figure}
\epsscale{1.0} \plotone{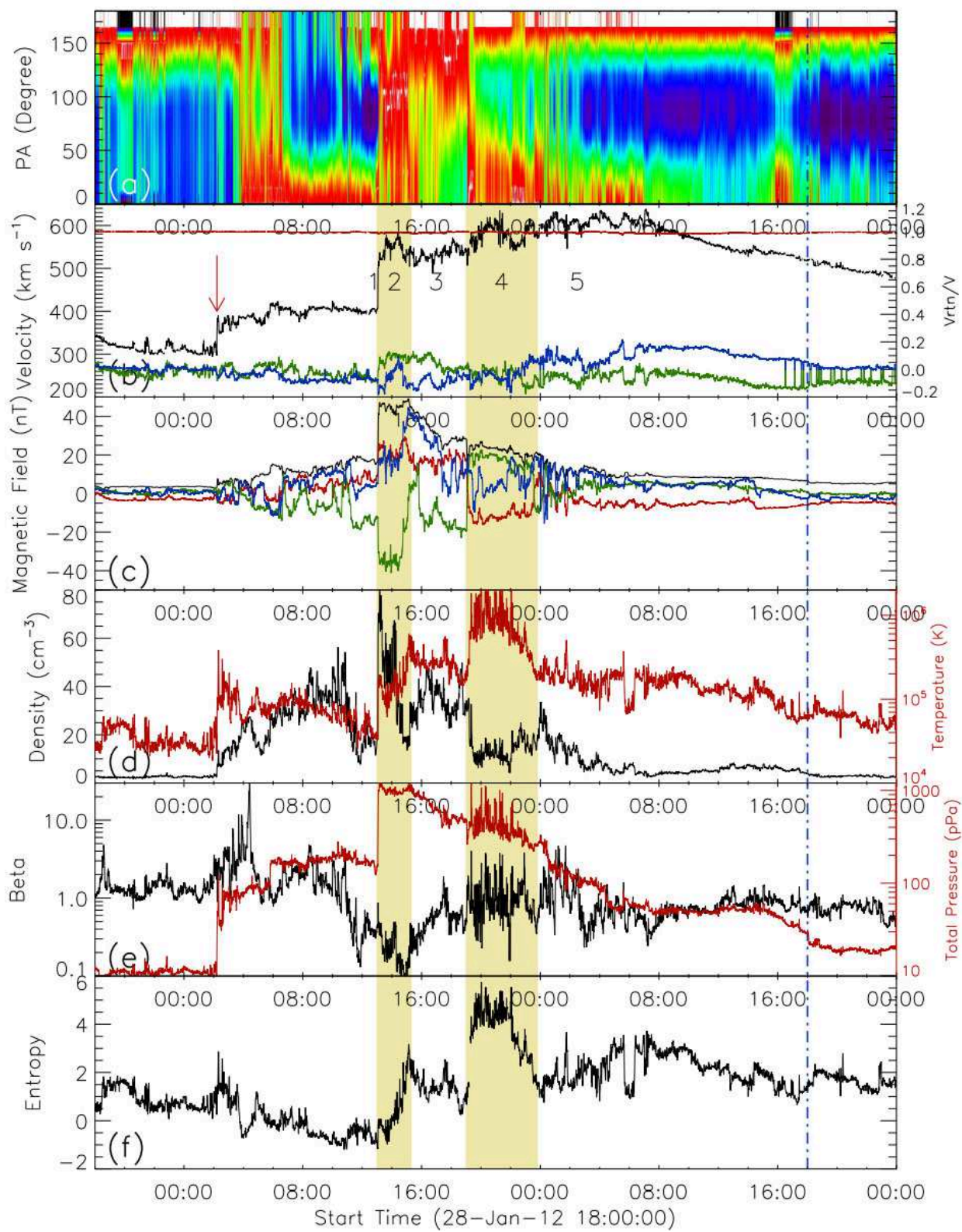} \caption{Solar wind parameters
measured with \textit{STEREO} A. From top to bottom, the panels
show the PA distribution of electron at 93.47 eV, bulk speed,
magnetic field, density and temperature, plasma $\beta$ and total
pressure, and entropy. See text for details. (A color version of
this figure is available in the online journal.)\label{fig5}}
\end{figure}

\end{document}